# High availability using virtualization

Federico Calzolari[1,2,3], Silvia Arezzini[2], Alberto Ciampa[2], Enrico Mazzoni[2], Andrea Domenici[3], Gigliola Vaglini[3]

[1] Scuola Normale Superiore - PISA ITALY
[2] National Institute of Nuclear Physics INFN - PISA ITALY
[3] University of Pisa, Department of Information Engineering (DIEIT) - PISA ITALY

E-mail: federico.calzolari@sns.it

**Abstract.** High availability has always been one of the main problems for a data center. Till now high availability was achieved by host per host redundancy, a highly expensive method in terms of hardware and human costs. A new approach to the problem can be offered by virtualization. Using virtualization, it is possible to achieve a redundancy system for all the services running on a data center. This new approach to high availability allows the running virtual machines to be distributed over a small number of servers, by exploiting the features of the virtualization layer: start, stop and move virtual machines between physical hosts. The 3RC system is based on a finite state machine, providing the possibility to restart each virtual machine over any physical host, or reinstall it from scratch. A complete infrastructure has been developed to install operating system and middleware in a few minutes. To virtualize the main servers of a data center, a new procedure has been developed to migrate physical to virtual hosts. The whole Grid data center SNS-PISA is running at the moment in virtual environment under the high availability system.

## 1. Introduction

One of the most critical issues for a computing center is the ability to provide a high availability service for all the main applications running on it. Today all the services are intended to be 24x7 (24 hours at day, 7 days per week), even those with a short life span.
Someone said "When the world can access your applications, application failures are exposed to a much wider community". High availability takes care of the strategies to reduce to the minimum the applications' downtime [1] [2] [3] [4] [5].
The strong demand for high-availability solutions has generated a lot of design strategies to provide a reliable service. These strategies involve systems such as heartbeat and cluster computing - very reliable but at the same time very expensive solutions in terms of hardware and human cost.
A lot of commercial solutions exist at the moment to satisfy the needs of service availability in a production environment. The problem is often related to the low portability of the solution, due to the strong dependency between the high availability tool provided and the operating system, or to the need of a specific software [6] [7] [8]. On top of that, a lot of proprietary solutions are not free, with an often too high cost for a research computing center or for a small company.
While high availability services are essential for 24x7 mission critical applications, the cost issue has to be carefully evaluated. To extend this service level to above a 99.9 percent (three nines) availability, the cost increases exponentially. Because of the high costs and hardware configuration requirements, a five nines availability level often implies a strong negative return on investment [9] [10] [11] [12] [13].

Our idea is to satisfy the research and the enterprise needs of high availability with a zero cost new solution. This new approach is based on the concept that a "relaxed" system may ensure the application redundancy required in the greater part of cases. "Relaxed" means that a system is able to restore any previously running application in less than ten minutes from the crash time.

3RC is the name of the project, acronym for 3 Re Cycle. The originality of this new approach to high availability is that a computing center system manager does not have to worry about the system redundancy till the disaster occurs. At that moment the system is able to restore the crashed application in a new location selected by a choice algorithm. With no additional costs in terms of hardware and software for a computing center, and slight operating costs, this new approach offers the ability to guarantee a relaxed high availability level. The whole Grid data center SNS-PISA is running at the moment in a virtual environment under this new high availability system.

A complex infrastructure, based solely on servers available in a typical computing center, has been developed in order to install the operating system and the middleware in a few minutes, by exploiting Preboot Execution Environment (PXE) technology, as well as DNS and DHCP services.

## 2. Scenario: Grid Data Center

INFN-PISA participates in the LHC experiments as a Tier2 regional center [14] [15] [16] [17] [18] [19] [20]. The computer center [21] consists of 2.000 CPUs, 500 TB disk and a full 1 Gb/s switching infrastructure, and it is expected to reach more than 5.000 CPUs and 1 PB of disk space in the next two years.

We decided to implement the Tier 2 as part of a common infrastructure between INFN, Scuola Normale, and Physics Department of Pisa University [22] [23] [24] [25] [26] [27] [28].

*2.1. System requirements*

In order to implement an architecture able to bootstrap a new host via network PXE, there are no other requirements than servers usually involved in the daily operation of a computing center, such as DHCP, TFTP, and HTTP servers [29] [30].

By modifying the DHCP configuration file, a string is passed to the host at boot time. Exploiting the PXE architecture, a complete system has been developed to automatically install operating system and middleware. The system has been successfully used for the installation of 2.000 CPUs progressing in parallel. Exploiting an almost unused field in DNS record, the HINFO (Host Info) field, in INFN Pisa we have set up a post install script able to perform all the needed operations to install middleware, patches and applications on all the computing center servers.

*2.2. Relaxed high availability model*

A primary idea on which the Grid infrastructure is based is that each node can be purchased as commodity hardware. A Grid computing center is therefore made out of low cost devices. Assuming that each central service is critical for the single computing center (not for the whole Grid), but it can bear a system down of 5 to 10 minutes, we introduced the new concept of "relaxed" high availability.

## 3. Storage

*3.1. Open issues in storage management*

Digitizing the information stored in every book in the Library of Congress requires about 10 terabytes of disk space. Storing all the movies ever made on DVDs needs 5 petabytes of data. Five petabytes are also the amount of data the Large Hadron Collider of CERN expects to collect annually in coming years. Half to one petabyte is the space required to store all the virtual machine disks of a medium (1.000 CPU) computing center.

Storing such large amounts of data in a single computing center requires a corresponding increase in aggregate bandwidth for the physical storage system, and the growth of I/O and RAM requirements for the disk servers providing access to the shared storage. Increasing the computational power and the storage capability in a data center by a factor N entails in many instances an increase of the aggregate bandwidth from/to the storage by a factor of up to $N^2$ [31]. Optimizing the storage layout can significantly help in the area of High Performance Computing and High Availability in virtual environment [32].

*3.2. Case studies and solutions*
A set of experimental data taken during the last five years at the Italian National Institute of Nuclear Physics (INFN) and Scuola Normale Pisa, several architectural storage solutions have been tried, and a lot of storage performance tests have been carried out [33] [34].
While Direct Attached Storage (DAS) can be useful for a small set of data shared from a single server, Network Attached Storage (NAS) and Storage Area Network (SAN) are the best choice for a large quantity of data to be shared over a large number of hosts. The use of NAS rather than SAN depends essentially on the available budget. At the moment a single NAS system can reach an I/O throughput of 4 Gb/s, while a complex - and very expensive - SAN can reach an I/O throughput of 100 Gb/s (and more). With respect to data redundancy and reliability, the chosen solution has been almost always the RAID 6 technology, due to the high capability of the single disks today. In case of a very limited budget, a Distributed Replicated Block Device (DRBD) has been developed to reach the same data reliability with no additional costs.

## 4. Proposals and solutions

*4.1. Background and motivations*
By analyzing the causes of planned and unplanned downtime of a large computing center (INFN Pisa: 2.000 CPU, 500 TB disk) for a period of three years, a new approach to high availability has been supplied, based on the features of virtual machine running on virtual environment [35].

*4.2. High availability using virtualization*
Since a not critical availability level is often required by almost all services running on web, Grid, or other technological environments, an interesting solution is offered by a new availability idea: a system able to restore an application in a few minutes. Usually a 2 to 3 minutes blackout is not critical for most part of the production applications. By paying for the non immediate recover of a crashed application, a new approach to high availability is offered by intensively exploiting all the features of a virtual environment.
While high availability services, such as cluster or heartbeat, are usually able to recover from a failure in a few seconds, a relaxed system is a solution able to restore a service in a few minutes.
A High availability virtual infrastructure ensures that a service has constant availability of network, processors, disks, memory. This way a failure of one of these components is transparent to the application, with a maximum time delay of two to ten minutes.

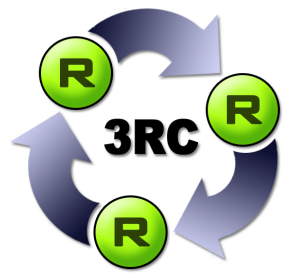

**Figure 1.** 3RC logo [by Claudio Atzeni]

Exploiting virtualization and the ability to install a host from scratch in a few minutes, it is possible to do a sort of host on-demand, where the start-up of a backup virtual machine is done only when the disaster occurs.

*4.3. Redundancy in virtual environments - 3RC*
The 3RC system offers several possible redundancy strategies, depending on the desired high availability level. There is not a unique solution for all possible requirements; each use case needs a different solution.
The system is able to recover all the managed services running in virtual environment in case of crash or failure. The unavailability of the original physical host providing the virtual machine - e.g. due to a physical machine fault - is managed in an absolutely transparent manner.

*4.4. Architecture and Infrastructure*
The solution has been developed in a virtual environment by using a remote controller that can be represented as a finite state machine. In each state an action can be performed on the single virtual machine hosting the crashed application or on the whole virtual layer running over the physical hosts.
A complex infrastructure, based solely on servers always available in a typical computing center, has been developed to install operating system and middleware in a few minutes.
The remote controller is the main core of the 3RC high availability system. It is responsible of checking and monitoring all the hosts involved in the high availability service.
The only requirement for the controller is the ability to access as administrator all the involved hosts.
The end choice for the virtualization layer was for to the VMware free server solution. The reasons for that choice were the VMware business reliability and the guarantee of a continuous software package update.
While the main controller of the high availability system needs a Linux based host, the only requirement for the servers hosting the virtual machines is to run the VMware server. The virtual machines can run whatever kind of operating system, from Unix to Windows.
Many monitoring systems have been tested to check the hosts status. Even if some of them are able to perform actions on the monitored machines, the chosen solution has been the Ganglia Monitoring System, an additional plugin has been developed to make Ganglia information available for a Linux shell script. It periodically analyzes the XML Ganglia log in order to manage the information in a more flexible and usable way.
A control script (the remote controller) is scheduled to run every 60 seconds on a single host - the host can be both internal or external to the cluster. 70 seconds is the average time needed from Ganglia to be aware of a machine operating system fault. The consequent average total time from a machine crash to its awareness by the 3RC controller is 30 + 70 seconds, with a confidence interval of ± 30 seconds. It means that 100 +/- 30 seconds are needed to detect a machine crash.

*4.5. Finite state machine approach*
The intervention level on a crashed machine, in the 3RC project, goes from a simple host reboot till a complete operating system and middleware re-installation, through a restart service.
A cyclic finite state machine has been developed to meet the required intervention levels. Sequentially switching through its three states, in a sort of intervention escalation, it operates in three different manners on the crashed system. The three intervention levels are: reboot, restart, reinstall.
The remote controller scans the whole cluster once per minute. In case of a failure detected, it checks the previous state of the virtual machines involved in the failure, and acts consequently. Only if the reboot or restart procedures have already been tried, the reinstall procedure starts.

*4.5.1. Reboot*
The first and simplest solution for a high availability service is to reboot the machine when an operating system crash occurs. This could be very difficult to achieve: the reason is that if an operating system crashes, with a large probability it is not reachable from outside. In case of machine isolation, the reboot cannot be induced from outside.

### 4.5.2. Restart

The term "restart" in this work means the restart of the virtual layer managing the crashed virtual machine. This way is it possible to switch a virtual machine on and off, as a real one. While a human intervention is needed to switch on a turned off physical host - with the exception of servers provided with remote hardware controller listening on the network - switching a virtual machine on can be done also by a remote controller.

By doing a restart of the virtualization layer managing the virtual machine, the system can easily operate a complete restart of the virtual machine, including the bootstrap procedure.

The restart of a virtual machine can be performed on the same physical host previously hosting the virtual machine, or on another physical host, depending on a choice algorithm based on:
- physical hosts availability;
- average load of physical hosts in the last 5 minutes;
- number of virtual machines already hosted by the single physical host.

A threshold load level is given for each physical server, depending on its CPUs number, RAM availability, internal processor architecture.

This way, if a physical host serving one or more virtual machines crashes, the hosted virtual machines can be easily moved to another physical host as soon as the controller detects the failure (the control check is scheduled once per minute). If no one of the available physical hosts in the cluster are at the moment free, the virtual machine waits to be restarted till a physical host reaches an under-threshold load level.

### 4.5.3. Reinstall

As a last chance, the system can provide a complete reinstall procedure, exploiting the whole PXE infrastructure previously described. After the reboot and/or the restart procedure have been performed unsuccessfully, the remote controller changes the link between the crashed virtual machine MAC address and the installation configuration required in the tftpboot/pxelinux.cfg path. The information about which operating system and middleware are required for a specific host are provided by a configuration file.

After the restart of the virtual machine, at bootstrap, the new kernel is loaded and the new installation starts from scratch.

The reinstall procedure can be disabled on a set of virtual machines, depending on the specific requirements - e.g. in case of non automatic installation provided for that kind of host.

As for the restart procedure, the reinstallation can be performed on the same or on another physical host, depending on the physical host availability, average load and status.

### 4.6. Recovery time

More than 15.000 crash test have been performed over 4 different testbed during a two week stress test period. Both the host and the guest operating system used are Linux: Debian the host, Red Hat the guest one.

### 4.6.1. Non destructive crash

For a simple non destructive crash the chosen procedures have been:
- simple switch off: the virtual machine is switched off by a halt, or shutdown command;
- high load increase: a recursive procedure aimed at increasing the virtual machine load until the system crashes or does not respond.

The time for a virtual machine Red Hat operating system Reboot is about 80 seconds [10 seconds for the PXE network setup procedure + 70 seconds for the boot] ± 10 seconds

The boot time added to the time the controller needs to recognize a failure gives the total recovery time in case of non destructive crash: 180 seconds ± 30 seconds.

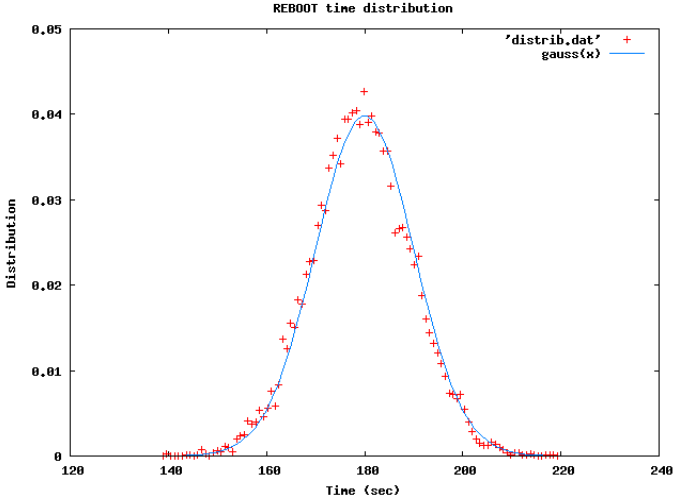

**Figure 2.** Recovery time distribution after 10.000 non destructive crash test

*4.6.2. Destructive crash*

For a destructive crash the chosen procedure used has been a reboot after the /boot partition has been erased. This is, in our opinion, the only safe way to certainly destroy a Linux based system.

The time for a virtual machine Red Hat operating system reinstall is about 442 [10 seconds for the PXE network setup procedure + 352 seconds for the installation + 80 seconds for the boot] ± 17 seconds.

The install time added to the time the controller needs to recognize a failure gives the total recovery time in case of destructive crash: 542 seconds ± 45 seconds.

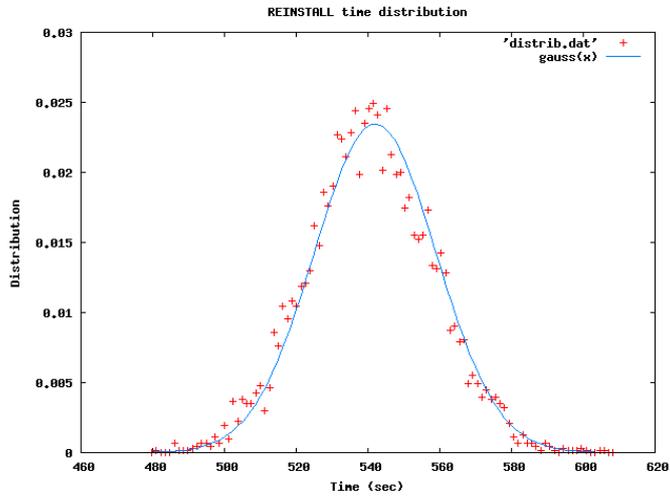

**Figure 3.** Reinstall time distribution after 5.000 destructive crash test

## 5. Operations

*5.1. Environment*

The 3RC high availability service is aimed at taking care of services running over a computing cluster, a set of physical hosts serving more virtual machines in virtual environment.
Ganglia is the monitor used to check the machine (both physical and virtual) health status. A redundant remote controller - the core of this project - attends to the system overview, and manages the actions to be taken in case of failure of one or more of the system components.

*5.2. Operational steps in case of host failure*
In this section we describe more in detail what happens when a physical or virtual host crashes.
When a host fails, the controller is able to detect the failure in less than one minute after the Ganglia monitor event detection.
As a consequence of a typical crash event - e.g. an overload of the virtual machine running a specific service - the monitor detects the failure 70 seconds after the failed host becomes unresponsive. The controller notices the failure with a delay of 0 to 60 seconds and starts the recovery procedure.
If the previous host state (the state of the virtual machine at the previous check, usually one minute before) was "up and running", the system try to do a reboot of the virtual machine involved in crash, hoping the machine is still reachable via ssh. After that, the system is in a sort of standby state, waiting till the end of T1 (3 minutes).
At the first check scheduled after T1, if the failed host is still unresponsive and no feedback comes back, the controller performs a restart of the virtual layer involving the failed virtual machine; this is equivalent to physically turn off and on the host, in order to restart operating system and all the loaded modules and applications.
If after a T2 (3 minutes) waiting time the host does not become responsive, compatibly with the system configuration permissions, the controller starts the complete reinstall procedure, performing a full installation from scratch: disk partition, disk formatting, operating system installation, post configuration process, middleware installation and configuration, services start up.

*5.2.1. Virtual / Physical machine failure*
When a virtual machine fails - e.g. in case of system crash, software overload, virtual device failure - the remote controller, after the detection - via monitor - of the host failure, performs one of the three possible steps in order to recover the host.
In case of physical machine failure - e.g. broken hardware - the system detects, based on a choice algorithm, the most appropriate host in the cluster to recover the hosted virtual machines. The virtual machines involved in the physical machine crash are moved to a backup physical host in the cluster.

*5.3. Operation in a real crash example*
At 4:00 AM (night local time) *gridce*, the computing element - the main server - of the SNS-PISA CERN LCG/EGEE Grid environment was turned off because of an electrical power glitch in the computing room (the causes of the electrical problem are unknown, as often happens).
At the time of the glitch the virtual machine *gridce* was hosted by the physical server *alfa01*. At that moment the *alfa01* load was higher than the maximum acceptable load (the information is stored in the hosts configuration file). The 3RC choice algorithm searched for another physical host in the computing cluster with a load level lower than threshold: *alfa04*. After an unsuccessful reboot try via ssh (the host is down), the virtual environment involving the crashed virtual machine *gridce* is restarted on *alfa04* physical host.
In less than eight minutes the service is completely restored. The time could have been reduced, by omitting the reboot step, to about 3 - 4 minutes.
The next figure shows the *gridce* crashed virtual machine, and the *alfa01* and *alfa04* physical hosts status during the 24 hours including the electrical power glitch.

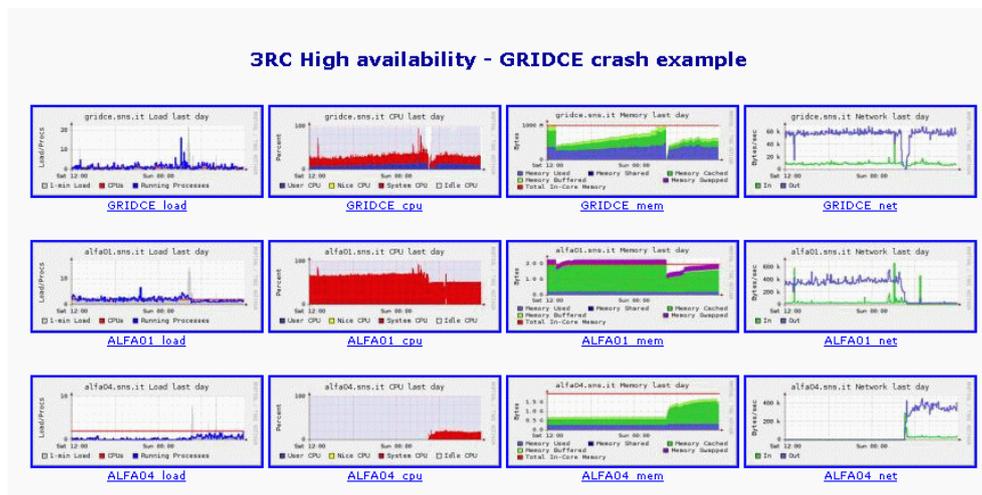

**Figure 4.** Status of the machines involved in the crash

## 6. Conclusions

By exploiting virtual environments and their features, a high availability service (3RC) has been developed with no additional costs in terms of hardware or software for a computing center, and a very low human effort. This new approach - compliant with the initial requirements - is able to guarantee a relaxed high availability level, with a recovery time of about three minutes for a non destructive crash, and lower than ten minutes if a compromised host needs to be completely reinstalled. The whole Grid data center SNS-PISA is running at the moment in virtual environment, controlled by the 3RC high availability system.

The main difference with respect to the pre-existing solutions is that no hardworking is required on the computing center hosts in order to guarantee the high availability service. On the other hand the 3RC high availability system does not provide a hot-swap host redundancy. It allows hosts to be restarted or reinstalled in a few minutes, but the operational continuity is not guaranteed.

It would be nice to refer to a statement by Luigi Picasso, Theoretical Physics Professor at the University of Pisa: "It is important to know what a theorem states, but it is probably more important to know what a theorem does not state". In line with this statement, what is this project not intended to be used for? The 3RC high availability system does not provide a zero downtime recovery service; it provide a relaxed recovery solution. Therefore, it is not a reliable solution for mission critical applications, such as financial transactions, security certificates management, real time controllers, human health related applications.